\documentclass{desyproc}

\begin{document}
\title{Terrestrial WIMP/Axion astronomy}

\author{{\slshape Ciaran A. J. O'Hare$^{1,2}$\\[1ex]
$^1$University of Nottingham, Nottingham, United Kingdom\\
$^2$Universidad de Zaragoza, Zaragoza, Spain}}

\contribID{ohare\_ciaran}

\confID{13889}  
\desyproc{DESY-PROC-2017-XX}
\acronym{Patras 2017} 
\doi  

\maketitle

\begin{abstract}
Predicting signals in direct dark matter (DM) detection experiments requires an understanding of the astrophysical structure of the local halo. Any uncertainty in this understanding will feed directly into all experimental results. However our terrestrial experiments are in a position to \emph{study} this same astrophysical dependence, and in fact represent our only probe of the local halo on sub-milliparsec scales. This is best achieved in the case of WIMP dark matter if directionally sensitive experiments are feasible, but requires novel parameterizations of the velocity distribution to make model independent claims. For axions the prospects are much greater, haloscopes would be able to make better measurements of the local DM distribution than astrometric probes.
\end{abstract}

\section{Introduction}
The relative motion of the Solar System with respect to the largely non-rotating DM halo of the Milky Way gives rise to an anisotropic flux of DM aligned with Galactic rotation. This peak direction is characteristic of a dark matter signal but moreover, the distribution of incoming velocities will encode the galactoarchaeological history of the halo. Measuring this and understanding it will undoubtedly give insight into the formation of the Milky Way and perhaps galaxy formation in general. For instance one may desire to accurately measure the velocity of the Solar system with respect to the halo, measure any underlying anisotropy or rotation in the DM halo, or uncover substructure components left over from the hierarchical construction of the halo by tidal accretion. Additionally in the case of axions there may be unique structures (unseen in vanilla CDM) associated with the primordial dynamics of the axion field. We discuss these prospects in the context of WIMP directional detectors (Sec.~\ref{sec:WIMPs}) and axion haloscope experiments (Sec.~\ref{sec:axions}), drawn from Refs.~\cite{Kavanagh:2016xfi, OHare:2017yze}, in which further discussion can be found.

\section{WIMP astronomy}\label{sec:WIMPs}
\begin{figure}[h]
\centering
\includegraphics[width=0.43\textwidth]{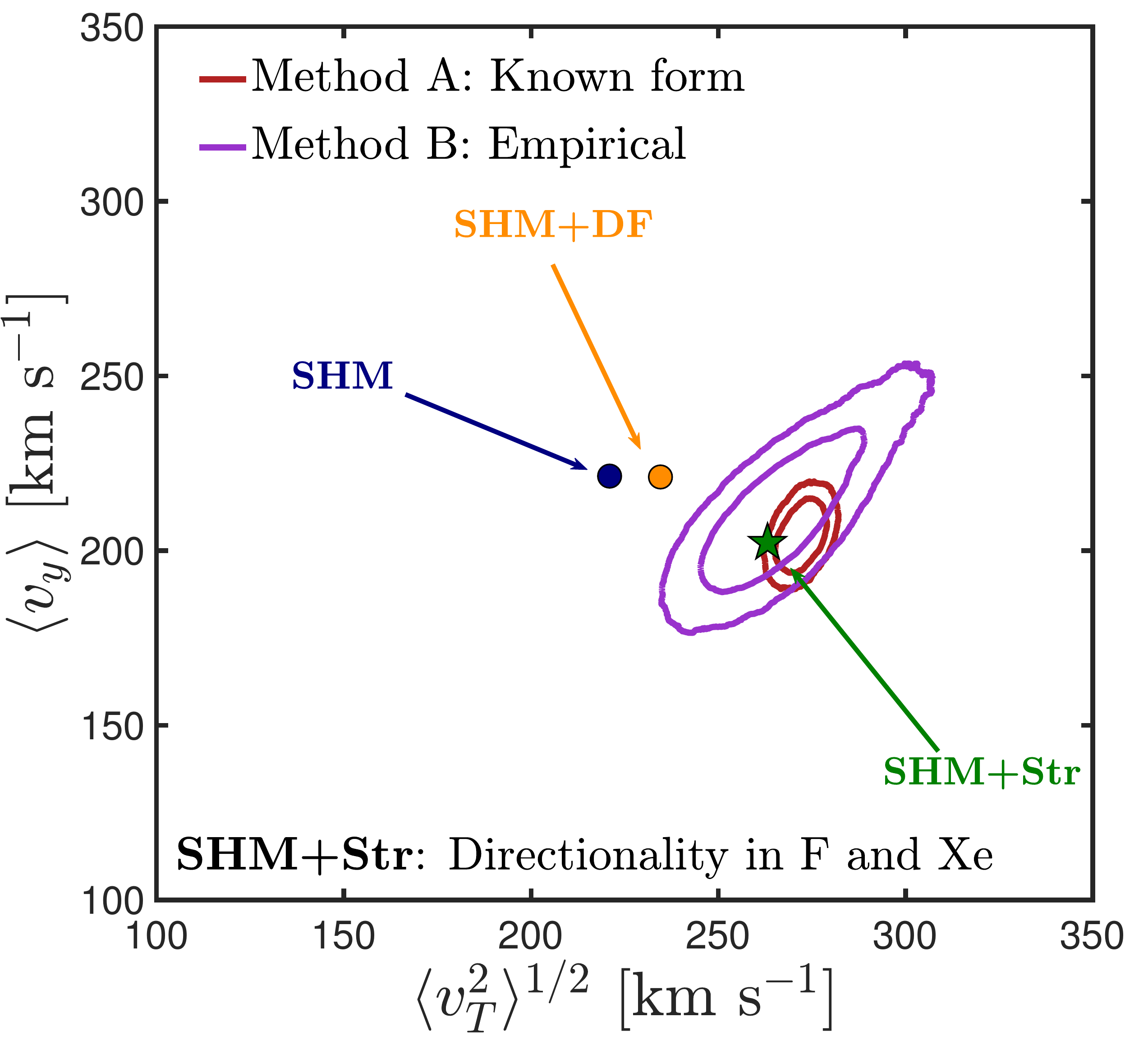}
\hspace{1.5em}
\includegraphics[width=0.43\textwidth]{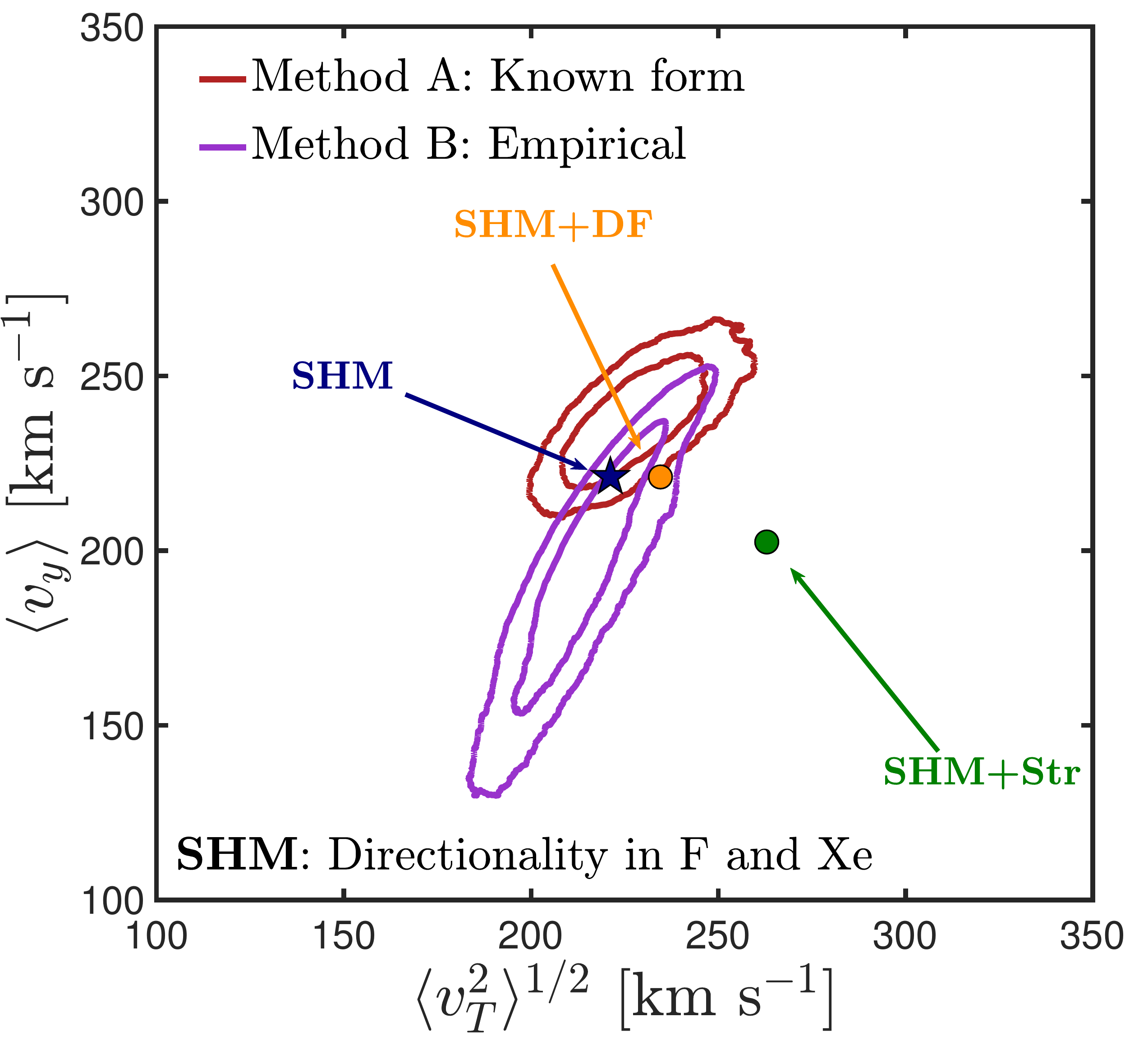}
\caption{68\% and 95\% confidence level contours for the reconstructed velocity distribution projected into the $\langle v_T^2 \rangle - \langle v_y \rangle$ plane (defined in Eq.(\ref{eq:vy})). We compare two methods of reconstruction A/B (described in the text). The markers indicate three benchmark models. The star in each panel indicates the correct underlying model. }\label{fig:vyvT}
\end{figure}
Probing the DM velocity distribution with WIMPs is possible to some extent if a sufficiently strong signal can be detected in some existing experiment. Unfortunately since conventional detectors measure only the energy of nuclear recoils, most of the 3-d velocity structure is lost (a small amount is preserved in the annual modulation). To fully extract this information in direct detection, one requires directionality (see Ref.~\cite{Mayet:2016zxu} for a review). Small-scale examples of such experiments already exist. The favored approach currently is to use low-pressure gas time projection chambers to measure mm-scale recoil tracks. Although these experiments are inherently low-mass, the technology is promising. Furthermore there is a possibility that a ton-scale nuclear recoil `observatory' will be constructed in the future.

The question is: how can we use directional information to reconstruct the velocity distribution, $f(\mathbf{v})$, in a model independent way? To describe our approach we directly compare the application of two methods on mock data:
\begin{itemize}
\item{{\bf Method A: Known form:} The functional form of $f(\mathbf{v})$ is known (but the parameters values are not.)}
\item{{\bf Method B: Empirical parameterization:} No knowledge is assumed about $f(\mathbf{v})$.}
\end{itemize}
We perform the empirical parameterization with a discretized approach. We discretize initially in three angular bins, $f^k(v)$, with $k=1,2,3$. Inside each bin there is a \emph{speed} distribution which is parameterized by a Chebyshev polynomial which can suitably capture the typical shape of a speed distribution while enforcing its normalization and positivity. We align the angular bins such that the $k=1$ bin points along the motion of the lab, $\mathbf{v}_{\rm lab}$, anticipating that the greatest anisotropy in the velocity distribution will be generated by the motion of the Earth through the halo. An advantage of this coarse-grained empirical approach is that it allows a completely model independent way of making claims about a given underlying velocity distribution. We show this by applying the method on data generated under three benchmark models that are inspired to cover the range of velocity structures possible in a real halo. These are 1) the standard halo model (SHM), 2) the SHM with the addition of a stream (SHM+Str), and 3) the SHM with the addition of a debris flow (SHM+DF).

We show in Ref.~\cite{Kavanagh:2016xfi} that the discretized approach is successful at capturing broad features in the velocity distribution, and especially at determining whether the velocity distribution contains excesses of particles in certain bins. An intuitive way of describing the success of the parameterization is to map the reconstruction onto physical parameters. We calculate mean values for the velocity parallel and transverse to the lab motion, respectively:
\begin{eqnarray}\label{eq:vy}
&\langle v_y \rangle& = \int \mathrm{d}v \,\int_{0}^{2\pi} \mathrm{d}\phi \, \int_{-1}^1 \mathrm{d}\cos\theta \, (v\cos\theta)\, v^2 f(\mathbf{v}) \, , \nonumber \\
&\langle v_T^2 \rangle& = \int \mathrm{d}v \,\int_{0}^{2\pi} \mathrm{d}\phi \, \int_{-1}^1 \mathrm{d}\cos\theta \, (v^2(1-\cos^2\theta))\, v^2 f(\mathbf{v}) \, .
\end{eqnarray}
In Fig.~\ref{fig:vyvT} we show two of the multiparameter reconstructions projected onto these parameters. The results show that the SHM and SHM+Str benchmarks can be distinguished above the 95\% level in Method B, where no prior knowledge is assumed. Results for a range of distributions and experimental setups are presented in Ref.~\cite{Kavanagh:2016xfi}.

\section{Axion astronomy}\label{sec:axions}
\begin{figure}[h]
\centering
\includegraphics[width=0.42\textwidth]{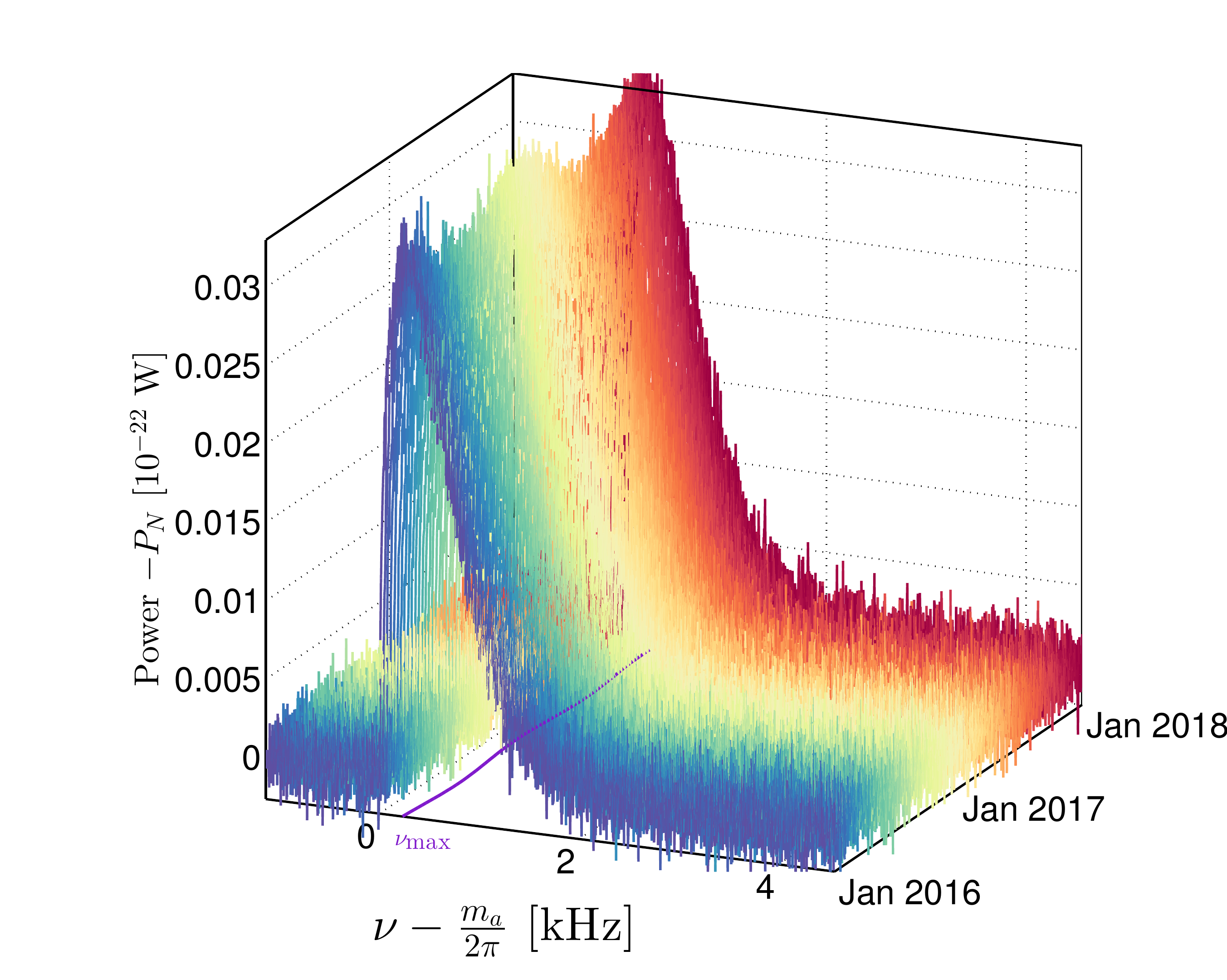}
\hspace{1.5em}
\includegraphics[width=0.42\textwidth]{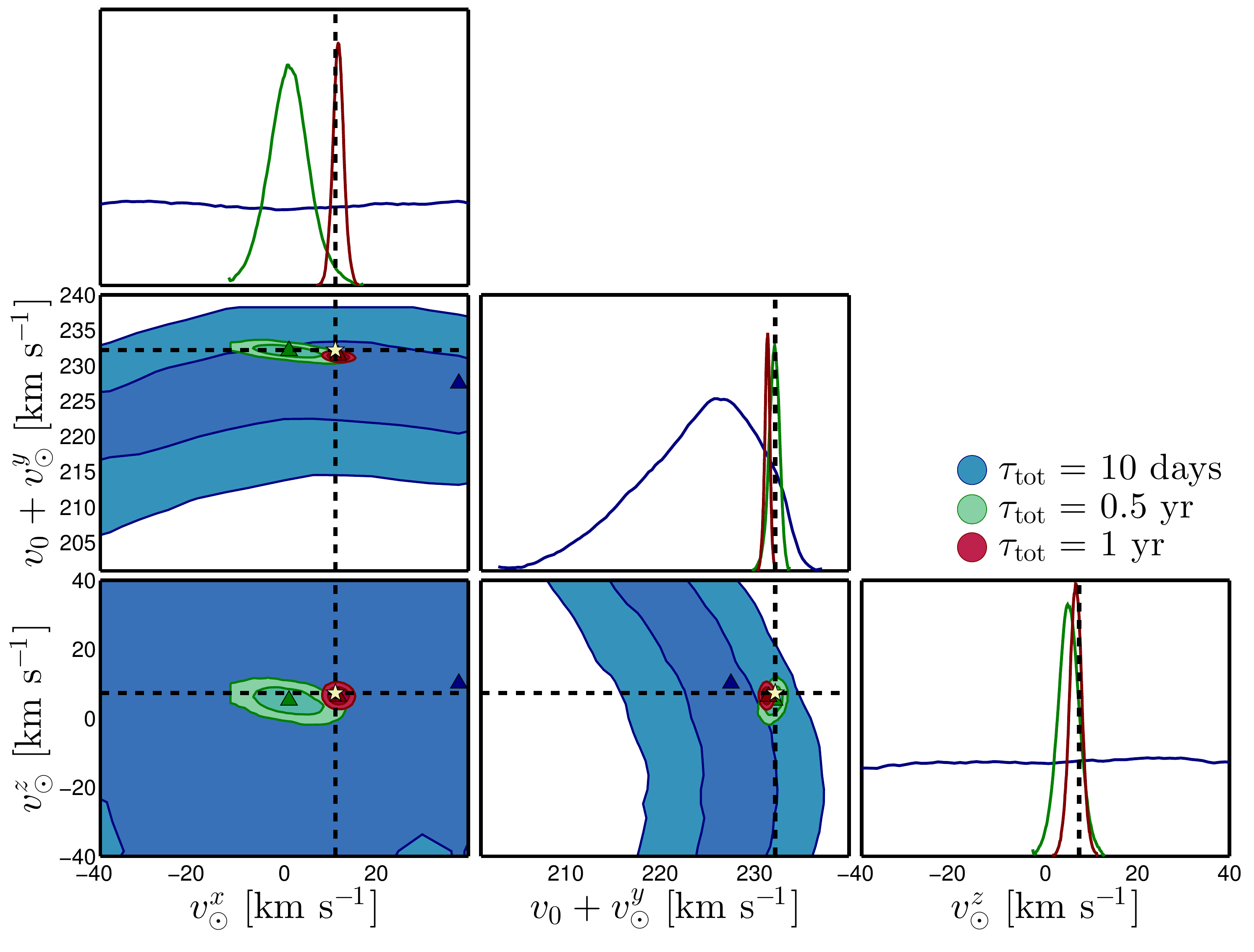}
\caption{{\bf Left:} simulated axion + noise power as a function of frequency (shifted by the axion mass) and time. The power displayed along the vertical axis is shifted by the mean noise $P_N$. {\bf Right:} reconstructed Galactic coordinate components of $\textbf{v}_\textrm{lab}$, displayed as 68\% and 95\% confidence level contours in the marginalized likelihood. Dashed lines show the input values of each parameter. The colors on each set of contours indicate the amount of data taking used to achieve the reconstruction, $\tau_{\rm tot}$. }\label{fig:axions}
\end{figure}
Experimental searches for DM axions are typically based on a coupling to electromagnetism that permits their conversion into photons inside a magnetic field. We follow a similar procedure to before, but inside a resonant microwave cavity experiment exploiting this process. In Fig.~\ref{fig:axions} we show a signal simulation of a hypothetical experiment based on ADMX that could be performed once the axion has been detected and a frequency range containing the axion mass has been identified. Like the nuclear recoil spectrum for WIMPs, the EM power received in a resonant cavity is dependent on the speed distribution $f(v)$. However since the axion is usually interpreted as an oscillating classical field, it is more natural to express this in terms of a power spectrum $|\mathcal{A}(\omega)|^2\propto f(v)$. The power spectrum is introduced by writing down the modes of the axion field $a(t) = a_0 \exp{(-i\omega t)}$, where $\omega = m_a(1 + v^2/2)$ (ignoring any spatial dependence). The spectrum of these Fourier modes is therefore related to the distribution of values of $v$, i.e. $f(v)$. Accounting for losses, the total signal power in a cylindrical cavity of volume $V$, quality factor $Q$, and magnetic field $B_0$, exploiting the first transverse electric field mode, is
\begin{equation}\label{eq:axionpower}
 P = g_{a\gamma\gamma}^2 B_0^2 V\omega_0 Q^3 \frac{4}{\chi_{0l}^2} \int_{-\infty}^{+\infty} \frac{\textrm{d}\omega}{2\pi} \mathcal{T}(\omega)|\mathcal{A}(\omega)|^2 \, ,
\end{equation}
where $\chi_{0l}$ is the $l$-th zero of the 0th Bessel function of the first kind. The cavity computation introduces $\mathcal{T}(\omega)$, which is a Lorentzian centered on the resonant frequency $\omega_0$, and describes the power loss off resonance. 

As before, we reconstruct sets of input astrophysics parameters. In Fig.~\ref{fig:axions} we show mock data comprised of multiple-day long time-integrated power spectra - collected over the course of one year to exploit the annual modulation due to the lab velocity. Here we display the reconstructed values of the Galactic coordinate components of $\mathbf{v}_\mathrm{lab}$. We show three sets of contours which correspond to experiments of different durations. The shortest 10 day long experiment corresponds to a single time-integrated bin of the 0.5 and 1 year long experiments. The phase and amplitude of the annual modulation is essential for measuring all three components, a feat nearly impossible to this accuracy in WIMP detection. The precision achievable with 1 year of data reaches the 1~km~s$^{-1}$ level, improving upon that of current astronomical observations~\cite{Schoenrich:2009bx}.

For axion astronomy we may also be interested in substructures that would not form from DM made of WIMPs. In particular, small bound structures of axions known as miniclusters are a consequence of the dynamics of the axion field at early cosmological times. These may be detectable on Earth today as we pass through the network of ministreams left behind as miniclusters become disrupted after passages through the stellar disk. These would show up prominently and characteristically as short lived features in the resolved power spectrum as we discuss further in Ref.~\cite{OHare:2017yze}.

\section{Summary}
We emphasize the differences between measuring a DM halo made of axions compared with one made of WIMPs. Even with an angular distribution of WIMP recoils, reconstructing and measuring properties of the local velocity distribution is difficult. However progress can be made with the use of empirical methods. However with axions, because we detect their conversion into photons - rather than via a stochastic scattering process - their kinematic structure is preserved and the prospects are much greater. This points towards the idea that in a post-axion discovery era haloscope experiments will be able to perform ``axion astronomy''. Finally, we remark that it may be possible to further push axion astronomy to directional sensitivity with the use of velocity dependent effects~\cite{Irastorza:2012jq, Millar:2017eoc}.

The author aknowledges support from the STFC and the University of Zaragoza.


\begin{footnotesize}

\end{footnotesize}


\end{document}